\documentclass{article}

\usepackage{latexsym}
\usepackage{graphicx}
\usepackage{color}
\usepackage{xspace}
\usepackage{soul}
\usepackage[bookmarks=false,linkcolor=blue,urlcolor=blue,colorlinks,citecolor=blue]{hyperref}

\usepackage[english]{babel}
\usepackage{authblk}
\usepackage[letterpaper,top=2cm,bottom=2cm,left=3cm,right=3cm,marginparwidth=1.75cm]{geometry}

\usepackage{amsmath,amssymb}
\usepackage{cite}
\usepackage{float}
\usepackage{xspace}

\newcommand{\ceq}[1] {(\ref{#1})}
\newcommand{\bisb}{${\rm Bi_{0.91}Sb_{0.09}}$\xspace}

\title{Phase-Slip Lines and Anomalous Josephson Effects in a Tungsten Clusters-Topological Insulator Microbridge}
\author[1]{Dong-Xia Qu}
\author[2,3]{Joseph J. Cuozzo}
\author[1]{Nick E. Teslich} 
\author[1]{Keith G. Ray}
\author[1]{Zurong Dai}
\author[1]{Tian T. Li}
\author[1]{George F. Chapline}
\author[1]{Jonathan L. DuBois}
\author[3]{Enrico Rossi}
\affil[1]{Lawrence Livermore National Laboratory, Livermore, CA 94550, USA}
\affil[2]{Sandia National Laboratories, Livermore, CA 94551, USA}
\affil[3]{Department of Physics, William $\&$ Mary, Williamsburg, VA 23187, USA}

\begin{document}
\maketitle

{\bf Superconducting topological systems formed by a strong 3D topological insulator (TI) in proximity to a conventional $s$-wave superconductor (SC) have been intensely studied as they may host Majorana zero modes. However, there are limited experimental realizations of TI-SC systems in which robust superconducting pairing is induced on the surface states of the TI and a topological superconducting state is established. 
Here, we fabricate a novel TI-SC system by depositing, via focused ion beam, tungsten (W) nanoscale clusters on the surface of 
TI Bi$_{0.91}$Sb$_{0.09}$. We find that the resulting heterostructure supports phase-slip lines that act as effective Josephson junctions.
We probe the response of the system to microwave radiation.
We find that for some ac frequencies, and powers, the resulting Shapiro steps' structure of the voltage-current characteristic
exhibits a missing first step and an unexpectedly wide second Shapiro step.
The theoretical analysis of the measurements shows that 
the unusual Shapiro response arises from the interplay between a static Josephson junction and a dynamic one,
and allows us to identify the conditions under which the 
missing first step can be attributed to the topological nature of the Josephson junctions formed by the phase-slip lines.
Our results suggest a new approach to induce superconductivity in a TI, a novel route to realizing highly-transparent topological Josephson junctions,
and show how the response of superconducting systems to microwave radiation can be used to infer the dynamics of phase-slip lines.} 
\medskip
\medskip

\noindent
{\bf Introduction}
\smallskip

\noindent
Hybrid structures 
formed by a strong topological insulator (TI) and a superconductor (SC)
have been theoretically predicted as a promising platform for realizing 
topological superconductivity~\cite{Fu2008, Sato2009, Hasan2010, Qi2011, Wang2012, Ren2019}. 
Soon after the theoretical proposals, experiments showed that superconducting pairing can
be induced on the surface states of three dimensional (3D) TIs~\cite{Veldhorst2012, Qu2012, Williams2012, Molenaar2014}.
Experimental studies of Josephson junctions (JJs) based on 2D or 3D TI-SC heterostructures then showed signatures in the current voltage
characteristic ($I$--$V$) under microwave radiation
consistent with the presence of a topological superconducting state~\cite{ Galletti2014, Wiedenmann2016,Charpentier2017}. 
Over the past few years, a growing number of JJs 
with 3D TI weak links have been realized and displayed signs suggesting the establishment
of a topological superconducting state~\cite{Schuffelgen2019, Rosenbach2021,Rosen2021, Bai2022}.
Recently, several studies have provided further insight into the behavior of JJs based on 
topological materials~\cite{Dartiailh2021,Rosen2021,Bassel2022,Cuozzo2022}, and, in particular,
have shown that signatures in the $I$--$V$ properties often associated with the topological
character of the superconducting state can also be observed in non-topological JJs~\cite{Dartiailh2021,Rosen2021,Bassel2022}.


%
%

The main challenges to realize a robust topological JJ based on heterostructures formed by a 3D TI and a SC are:
(i)   realization of an almost ideal TI-SC interface;
(ii)  suppression of disorder;
(iii) fabrication of short and very narrow JJs.
In this work, to overcome these challenges we follow a very different approach from previous ones:
to create the TI-SC heterostructure we deposit tungsten (W) clusters on TI Bi$_{0.91}$Sb$_{0.09}$ using the focused ion-beam technique (FIOB),
and to form the JJ we rely on the natural formation of phase-slip lines (PSLs), lines across which the phase
of the superconducting order parameter increases at different rates.
Forming the TI-SC hybrid system by deposing W clusters has two advantages: the W clusters, being separated and randomly placed, do not significantly
modify the electronic structure of the TI, and yet, can induce via the proximity effect pairing correlations in the TI's surface states at low temperature, given that the inter-cluster distance is comparable to the normal-metal coherence length of Bi$_{0.91}$Sb$_{0.09}$; 
it minimizes the exposure of the TI's surface to air and it removes
the need to perform any annealing, both of which can strongly affect the TI's surface properties and doping.
By relying on the natural formation of a PSL we can realize an effective JJ with a length of just few nanometers and a width 
controlled by the W coverage of the TI. Given that W is deposited via FIOB the JJ width can be as small as few 10s nm.

We find that the W clusters induce on 
Bi$_{0.91}$ Sb$_{0.09}$'s 
surface a superconducting state with a critical temperature $T_c$ that is slightly below the $T_c$ of W nanoclusters.
Transport measurements in the dc regime reveal that the system undergoes a Berenziskii-Kosterlitz-Thouless (BKT) transition. 
Jumps in the voltage-current ($V$--$I$) characteristic can be associated to the presence of phase-slip lines which form effective JJs. 
To probe the properties of such JJs
we measure the $V$--$I$ characteristic under microwave radiation for different ac frequencies and powers.
We find that at intermediate frequencies and powers the first Shapiro step is missing, and that at
low frequencies and powers, in addition to the first Shapiro step being missing, the second step can be very wide.
We develop the theory to explain such unusual features and find that for intermediate frequencies and powers
the missing step can be explained by the presence of Landau-Zener transitions (LZTs),
and that for low frequencies and powers the structure of the Shapiro steps 
can be understood considering the presence of two JJs, formed by PSLs, one of which has its effective width
dynamically driven by the biasing current. 
The results have important implications for achieving proximity-induced superconductivity in a TI, understanding 
how seemingly $4\pi$-periodic Andreev bound states (ABSs) might arise in Josephson junctions formed by PSLs,
and understanding how signatures of the ac response can be used to infer the dynamics of PSLs 
and the effect on such dynamics of the biasing currents.

\medskip
\medskip
\noindent
{\bf Results}
\smallskip

\begin{figure}[ht] \vspace{0pt}
\includegraphics[width=5.5 in]{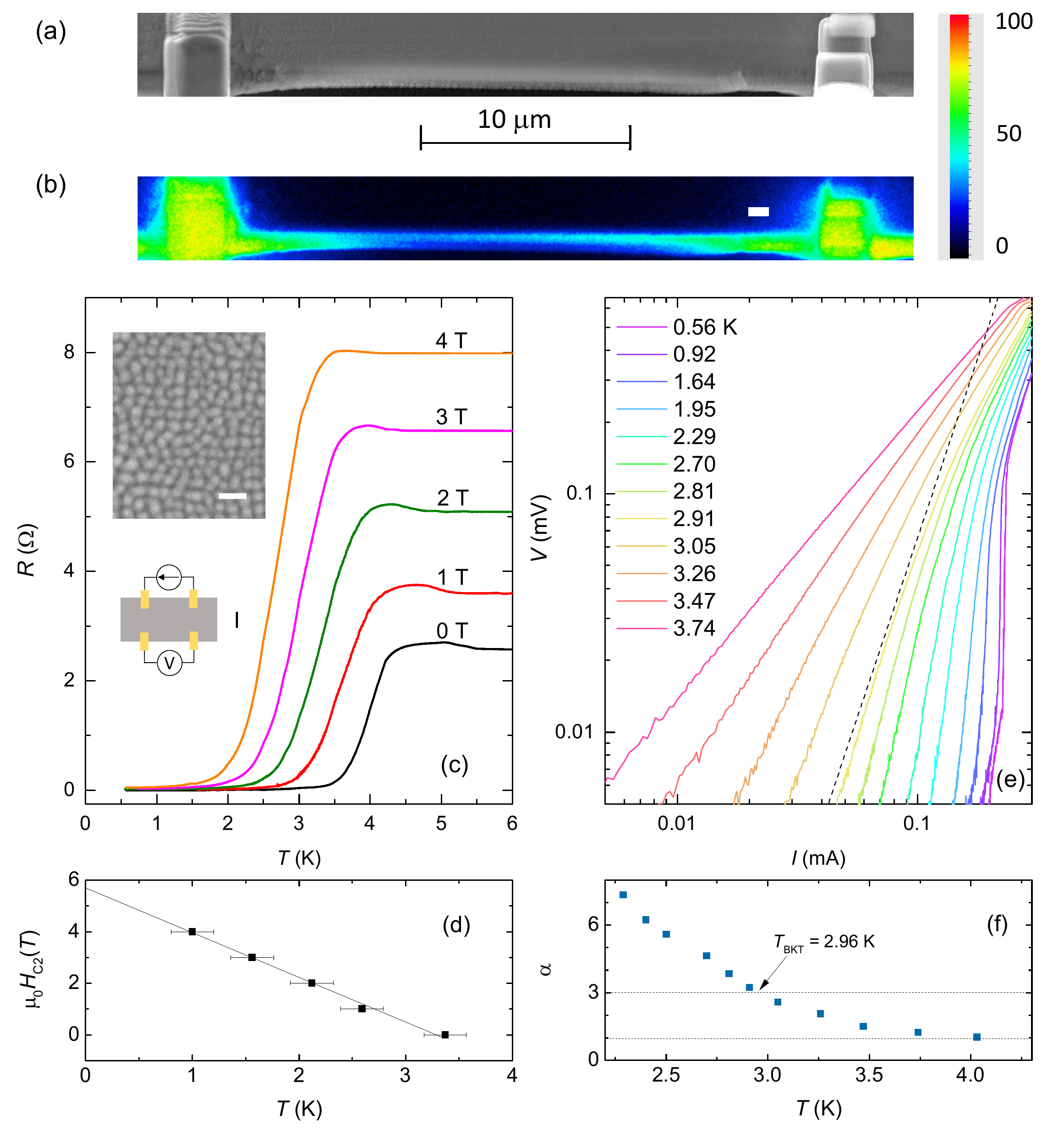} \centering
\vspace{10pt}\caption{\textbf{a}, Scanning-electron-microscopy (SEM) image of the sample, where superconducting W pads are fabricated on the Bi$_{0.09}$Sb$_{0.91}$ flake with a distance of $L \sim 30$ $\mu$m apart. Scale bar = 10 $\mu$m. \textbf{b}, The corresponding false-color energy-dispersive X-ray spectroscopy (EDS) elemental map shows the distribution of elemental W. The W clusters spread out around the W leads, forming a bow-tie shaped $\sim$1 $\mu$m by 30 $\mu$m microbridge. Scale bar = 1 $\mu$m. \textbf{c}, Resistance $R$ as a function of temperature $T$ for the 2.6-$\mu$m-thick sample measured using the probe configuration I (see bottom inset). The magnetic field is applied perpendicular to the sample surface and the bias current is 10 $\mu$A. Top inset: SEM image of W islands on the Bi$_{0.91}$Sb$_{0.09}$ substrate, taken at a distance of $2.8$ $\mu$m from a 200-nm-thick W deposit (scale bar = 200 nm). 
\textbf{d}, Temperature dependence of the upper critical field $H_{c2}$, which follows the GL theory for a 2D superconductor: $H_{c2}=\frac{\Phi_0}{2\pi \xi_{GL}(0)^2}(1-\frac{T}{T_c})$, where $\Phi_0$ is the flux quantum. \textbf{e}, $V(I)$ curves on a logarithmic scale. The long dashed line corresponds to $V$$\sim$$I^3$ dependence. \textbf{f}, Temperature dependence of the power-law exponent $\alpha$. The data $\alpha$ is extracted from the fits to the $V(I)$ curves shown in \textbf{e}.}\label{Fig1} \vspace{-10pt} \end{figure}

\noindent
We present results for devices in which W leads are grown using the focused-ion-beam technique on 
Bi$_{0.91}$Sb$_{0.09}$ flakes with a thickness of 2--5 $\mu$m.
%
%
Due to the halo effect \cite{Qu2013, Qu2018}, self-assembled W islands with a thickness of 10--50 nm form around the deposited W. Details about the fabrication and characterization of the devices can be found 
in the Methods section and Supplementary Information (SI).
We have studied the sample with the geometry shown in Figs.~\ref{Fig1} (a) and (b), 
in which a bow-tie-like strip of W islands was deposited within a 1-$\mu$m-wide region from the edge of the Bi$_{0.91}$Sb$_{0.09}$ flake to produce a microbridge.
The inset of Fig.~\ref{Fig1}~(c) shows a scanning-electron-microscopy (SEM) image of the W islands. We find that the island diameter is typically in the range of $50$--$60$ nm, and edge-to-edge spacing between islands is  $~20$ nm. 
The island size and inter-island spacing depend on the ion dose and gradually decrease with increasing distance from the deposition region.

We first perform dc measurements to characterize the superconducting state of the W-TI heterostructure.
The inset of Fig.~\ref{Fig1}~(c) shows the contacts' configuration used to measure the $I$--$V$ characteristic. 
Figure \ref{Fig1}~(c) shows the resistance $R$ versus temperature $T$ profiles under a perpendicular magnetic field, $H$, stepping from 0 to 4 Tesla. The normal-state resistance displays an upturn at low temperatures for all magnetic fields. This behavior arises from the current redistribution related to sample non-homogeneity together with an out-of-line contact arrangement \cite{Vaglio1993}. 
For $H=0$, at $ T \sim 4$ K, the system undergoes a broad superconducting transition, signaled by a sharp reduction of the resistance, while inter-island phase coherence develops \cite{Eley2012}. On further decreasing $T$ below $1.6$ K, the resistance vanishes completely and the global phase coherence is reached. 
Increasing $H$ decreases the temperature at which coherent superconducting states are established. Figure~\ref{Fig1}~(d) shows the value of the upper critical field $H_{\mathrm{c2}}(T)$ as a function of temperature.
%
A linear fit of this data allows us to estimate the in-plane Ginzburg–Landau (GL) 
coherence length at zero temperature to be $\xi_{GL}(0) = 7.6 \pm 1$~nm.
This value agrees with tungsten's superconducting coherence length, $\xi_W$.

Figure~\ref{Fig1}~(e) shows, on a logarithmic scale, the dc $V$--$I$ characteristic for $H=0$ 
and different values of $T<4$~K.
We see that when the current is larger than threshold values, that depend on $T$,
$V$ grows with $I$ following a power law, $V \propto I^{\alpha(T)}$, with a $T$-dependent $\alpha$. 
This indicates the presence of dissipation due to the motion of vortices and antivortices in the superconductor.
As $T$ grows the 2D superconductor undergoes a BKT transition 
at the BKT transition temperature, $T_{\mathrm{BKT}}$. For $T=T_{\mathrm{BKT}}$ 
vortex-antivortex pairs break and $\alpha(T_{\mathrm{BKT}})=3$ \cite{Kosterlitz1973, Halperin1979, Resnick1981, Chapline2008}.
The black dashed line in Fig.~\ref{Fig1}~(e) shows the slope, on the log-log scale, corresponding to $\alpha=3$. Figure~\ref{Fig1}(f) shows the evolution of $\alpha$ with $T$. We determine $T_{\mathrm{BKT}} = 2.96$~K from where $\alpha=3$ interpolates.
%



\begin{figure*} \vspace{0pt}
\includegraphics[width=6 in]{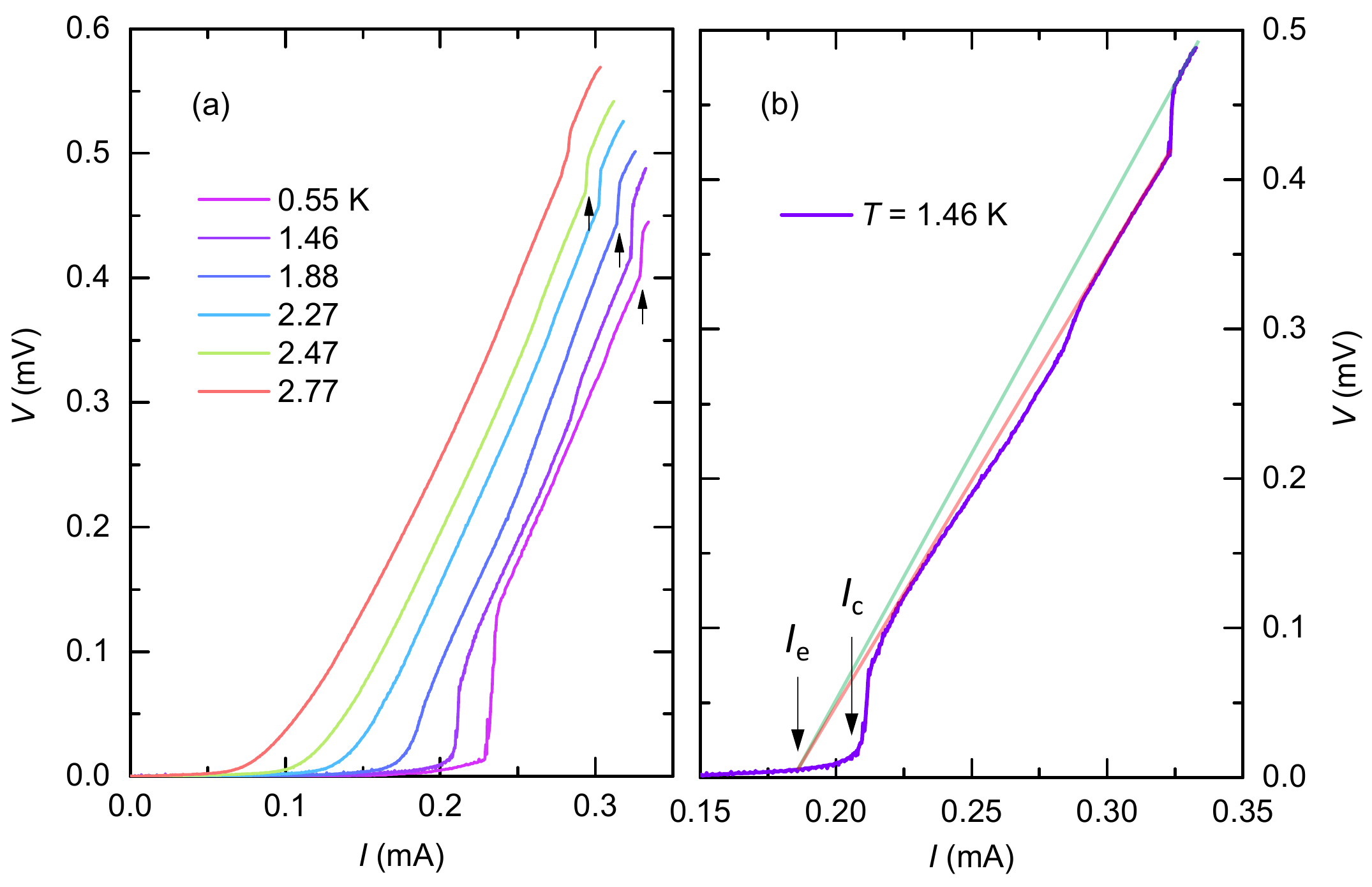} 
\vspace{-0pt}\caption{ \textbf{a}, Temperature dependence of $V$--$I$ characteristic obtained with configuration I. The black arrows indicate the second voltage jump at a higher current.  \textbf{b}, Voltage--current characteristic obtained with configuration I at $T = 1.46$ K. The red and green lines are extrapolated linear $V$--$I$ segments from the first and second resistive branches, respectively. These two resistive branches exhibit approximately the same excess current $I_e$, determined by the intersection of the red or green lines with the current axis. This behavior is consistent with the signatures of phase-slip lines previously observed in quasi-two-dimensional superconducting strips.}\label{Fig2} \vspace{-10pt} \end{figure*}

The results presented in Fig.~\ref{Fig1} show that our W-TI heterostructure is a proximity-coupled superconducting system \cite{Eley2012,Sun2018}.
By examining the $V$-$I$ characteristic at higher currents we observe the presence of additional voltage jumps
for $I > 0.25$ mA for all temperatures, Fig. \ref{Fig2}~(a). %
We find that the slopes of the $V$-$I$ characteristic before and after each additional jump approximately extrapolate at $V=0$
to the same current value, the so called excess current $I_e$, as shown in Fig. \ref{Fig2}~(b). 
The features of the dc $V$--$I$ characteristic at high currents 
are consistent with the formation of PSLs, resistive states arising in thin superconducting films when the current is larger than
a threshold value, 
$I_t$~\cite{Skocpol1974, Volotskaya1981, Dmitrenko1996, Androvov1993, Sivakov2003, Dmitriev2005, Berdiyorov2014}.
%
%
%
%
A PSL has width $\sim\xi$, the superconducting coherence length. In our case $\xi=\xi_W$ given that
\bisb's superconducting correlations are only induced by W via the proximity effect.
Across the PSL a voltage $V=R_{PSL}(I-\bar I_s)$ is established, where $I$ is the biasing current, $R_{PSL}$
is the effective resistance of the PSL, 
and $\bar I_s$ the average supercurrent across the PSL.
$\bar I_s$ can be identified with the excess current 
$I_e$, i.e., the current that crosses the PSL even when $V=0$.
As a consequence a PSL can be described effectively as a biased JJ, of length $\xi$, with critical current $I_c=I_e$.
%
The dependence of $dV/dI$ on the perpendicular field $B_\perp$ and dc bias current shows signatures
of a Fraunhofer pattern consistent with a JJ of length $L\approx\xi_W$.
Using an induced gap on Bi$_{0.91}$Sb$_{0.09}$ equal to W's superconducting gap, for all the Fermi pockets of \bisb's surface states, 
we obtain a coherence length 
that is at least a few times larger than $\xi_W$.
As a result,  
for Bi$_{0.91}$Sb$_{0.09}$’s surface states 
a PSL in W-TI hybrid can be well approximated as a short JJ.

For a superconducting TI, the effective JJ associated with a PSL can be expected to have a topological character.
In the presence of microwave radiation the $V$--$I$ characteristic of a JJ exhibits Shapiro steps~\cite{Shapiro1963} for $V= n hf/2e$, where
$f$ is the frequency of the radiation and $n$ is an integer.
For a topological JJ the current-phase relation (CPR) is $4\pi$-periodic~\cite{Kitaev2001, Kwon2004}
and this results in missing Shapiro steps for odd $n$~\cite{Wiedenmann2016, Bocquillon2017, Deacon2017, Rokhinson2012, Laroche2019}. 
However, in highly transparent JJs, Landau-Zener processes can cause the odd Shapiro steps to
be missing even when the junction is not topological~\cite{Dartiailh2021}.

\begin{figure}[ht] \hspace{-10pt}
\includegraphics[width=6.5 in]{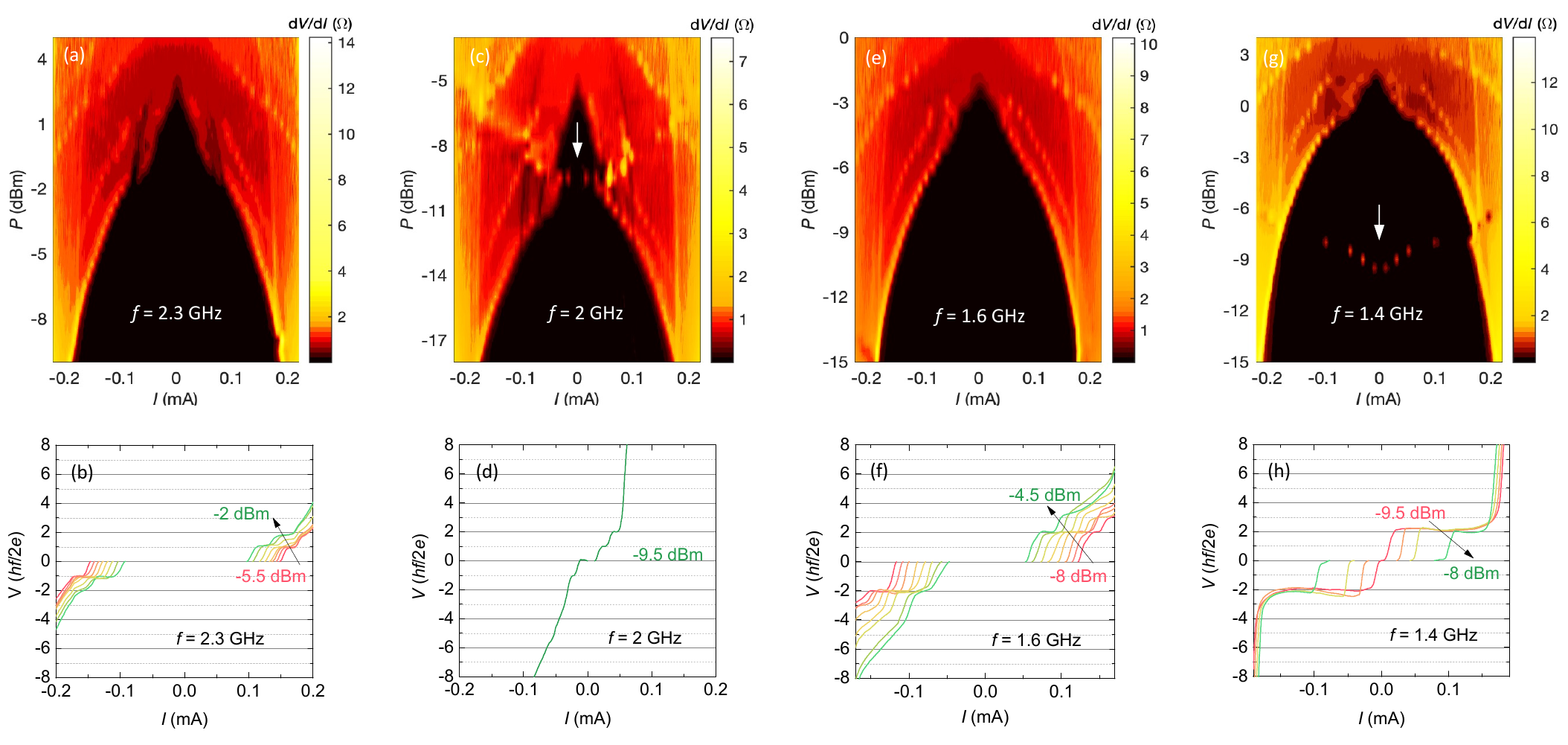} 
\vspace{-10pt}\caption{The ac Josephson effect measured using probe configuration I. 
\textbf{a}, \textbf{c}, \textbf{e}, \textbf{g}, color maps of the differential resistance d$V$/d$I$ as a function of the rf power $P$ 
and dc bias current $I$ for rf frequencies $f = 2.3$, $2$, $1.6$, and $1.4$ GHz at $T =0.56$ K. 
The white arrows in \textbf{c},~{\bf e} indicate the in-gap Shapiro response. 
\textbf{b}, \textbf{d}, \textbf{f}, \textbf{h}, Shapiro steps at different irradiation powers. 
The voltage is scaled in the unit of Shapiro voltage $\Delta V=hf/2e$.}\label{Fig3} \vspace{150pt} \end{figure}

Figures \ref{Fig3}~(a),~(c),~(e), and (g) show the color maps for d$V$/d$I$ versus the ac power $P$ and the bias dc current $I$ at microwave frequencies 
$f=2.3$, $2.0$, $1.6$, and $1.4$ GHz, respectively.  The corresponding $V(I)$ dependence, 
obtained from the integration of the d$V$/d$I$ curve over the peak area,
is shown in  Figs.~\ref{Fig3}~(b),~(d),~(f), and (h), respectively. 
At high frequency, $f=2.3$ GHz, we observe the usual structure for the Shapiro steps
consistent with a conventional $2\pi$-periodic CPR.
As $f$ is decreased, $f=2.0$~GHz, we observe the appearance of additional peaks in the $dV/dI$ at low bias currents that result
in regular Shapiro steps.  
As the $f$ is decreased further, $f=1.6$ GHz, we observe the disappearance
of the first, odd, Shapiro step indicating that the CPR of the JJ formed by the PSL has
a non-negligible $4\pi$-periodic component either due to its topological character~\cite{Wiedenmann2016, Schuffelgen2019, Dominguez2017, Bocquillon2017} or due to
Landau-Zener processes~\cite{Dartiailh2021}. 
Because no hysteresis is observed in our devices the missing steps cannot be attributed to hysteretic effects.
At even lower frequencies, $f=1.4$ GHz, the peaks in the $dV/dI$ at low bias currents result in a
Shapiro steps' structure in which the first step is absent, and the second one is unusually long. 
For the steps at low bias currents shown in Fig.~\ref{Fig3}~(h) we also notice that 
the in-gap critical current in the presence of an ac bias, $I_{c,ac}$ 
appears to increase with power, rather than decreasing, as in conventional JJs.
This suggests that in our system some properties, such as the width of the effective JJs created by PSLs, might 
be affected by the biasing current and ac power.

\medskip
\medskip
\noindent
{\bf Theoretical analysis}
\smallskip

\noindent
To understand the anomalous structure of the Shapiro steps shown in Fig.~\ref{Fig3}, we developed and studied 
an effective model to describe the JJs created by the PSLs. 
A calculation of the Shapiro steps from a microscopic model is computationally prohibitive for the size of our devices~\cite{rossignol_role_2019}, 
and so we describe the dynamics of the JJs using a resistively and capacitively shunted junction (RCSJ) model.
%
Within the RCSJ model, 
for a current-drive junction 
the dynamics of the phase $\phi$ across the junction  is given by:
\begin{align}
    \frac{d^2 \phi}{dt^2} + \sigma \frac{d\phi}{dt} + \frac{I_s(\phi)}{I_c} = \frac{I_{dc}}{I_c} + \frac{I_{ac}}{I_c}\sin(\omega t)
    \label{eq:JJ}
\end{align}
where $t=\sqrt{\frac{2eI_c}{\hbar C}} t^{\prime}$ is a dimensionless time variable, 
$\sigma = \sqrt{\frac{\hbar}{2eI_c R_n^2 C}}$ is the Stewart-McCumber parameter,
$I_s(\phi)$ is the supercurrent across the JJ,
and  $I_{dc}$, $I_{ac}$ are the dc and ac bias currents, respectively.
For $\sigma\gg 1$ the JJ is overdamped and we can neglect the first term on the left hand side of Eq.~\ceq{eq:JJ}
and simplify the model to a resistively shunted junction (RSJ) model.
From the dc transport measurements, Fig.~2, we extract $R_N\approx 8.4$ $\Omega$, and from
experimental results like the ones shown in Fig.~3~(a) we extract $I_c\sim 0.1$~mA.
Assuming $C\approx 1$~fF, the expected value for a JJ with a geometry similar to the JJ
formed by a PSL in our devices, we obtain $\sigma\approx 20$ (see SI). This implies that to 
understand the results shown in Figs.~3~(a),~(b), and Figs.~3~(e),~(f), to good approximation, we can treat the JJs as overdamped.
%
%

In general, for JJs based on superconducting TIs, we have that $I_s$ has both a $2\pi$-periodic, $I_{2\pi}$, component and $4\pi$-periodic one, $I_{4\pi}$.
%
Because the topological nature of the JJ only guarantees one crossing in the ABS's spectrum at $\phi=\pi$, it only contributes
one $4\pi$ mode to the total supercurrent across the JJ.
The maximum supercurrent $I_{c}^{(i)}$ carried by a single conducting mode is given by $I_{c}^{(i)}=e \Delta/2\hbar$. 
From the value of $T_c$ for W, $T_c=4.4$~K, we obtan $\Delta = 1.76 k_B T_c=668$ $\mu$eV 
and therefore $I_{c}^{(i)}\approx 81$ nA.
A junction with an $I_{4\pi}$ component exhibits missing odd Shapiro steps
for frequencies smaller than $f_{4\pi}=2eR_N I_{4\pi}/h$~\cite{Dominguez2012}.
As a consequence, if there is only one mode contributing to $I_{4\pi}$, we obtain $f_{4\pi}< 0.5$~GHz.
Given that we observe missing odd steps for $f>1$~GHz we conclude that to 
explain $dV/dI$ profiles like the one shown in Fig.~\ref{Fig3}~(e) (no in-gap steps)
we need to have more than a single mode contributing to $I_{4\pi}$.
%
%
Given the large width, $W > \xi$, of the bow-tie-like strip of tungsten islands, and therefore of the JJs formed by PSLs
located away from the center of the bow-tie,
we can have  Andreev mid-gap states with 
small gaps at $\phi = \pi$, 
and sizable detachment gaps from the continuum at $\phi = 0$ \cite{Dartiailh2021}. 
Such modes can contribute to the $4\pi$-periodic component of the supercurrent $I_s(\phi)$ given 
that they have a large probability, $P_{LZT,\tilde\tau}$,
to undergo a Landau-Zener transition (LZT) at $\phi = \pi$, and a negligible probability 
to undergo transitions at $\phi=0$ mod $2\pi$ into the continuum.
To good approximation we have \cite{Averin1995}:
\begin{align}
    P_{LZT,\tilde\tau}(t=t_{n\pi}) = \exp \left(-\pi \frac{\Delta (1-\tilde\tau)}{e \vert V(t_{n\pi}) \vert } \right),
    \label{eq:lzt}
\end{align}
where $t_{n\pi}$ is the time when $\phi\to(2n+1)\pi$ ($n\in\mathbb{N}$), 
$\tilde\tau$ is the average transparency of high transparency modes which also have a sizable detachment gap \cite{Dartiailh2021}, 
and $V(t_{n\pi}) = (\hbar/2e) (d\phi/dt)\vert_{t=t_{n\pi}}$. 
$dV/dI$ profiles like the one shown in Fig.~\ref{Fig3}~(e) can be understood considering an effective RSJ model
in which the supercurrent $I_s(\phi)$ has two channels~\cite{Dartiailh2021}:
one low-transparency channel with a purely $2\pi$-periodic CPR, 
$I_{s,2\pi}=I_{2\pi}\sin(\phi)$, and for which no LZTs can take place, 
and a high-transparency channel with 
$I_{s,\tilde\tau} =I_{\tilde\tau} \sin(\phi)/[{1-\tilde\tau\sin^2(\phi/2)}]^{1/2}$. 
To obtain the dynamics of the JJ we integrate 
Eq.~\ceq{eq:JJ}, neglecting the first term on the left hand side, setting
$I_s(\phi)=I_{2\pi} \sin(\phi) + I_{s,\tilde\tau}(\phi)$,
evaluating $P_{LZT,\tilde\tau}$ at times $t=t_{n\pi}$ and switching the sign in front of $I_{s,\tilde\tau}$ 
for $t=t_{n\pi}$ if a randomly generated number $0<r<1$ is smaller than $P_{LZT,\tilde\tau}(t_{n\pi})$. 

\begin{figure}[ht] \hspace{-10pt}
\includegraphics[width=6.5 in]{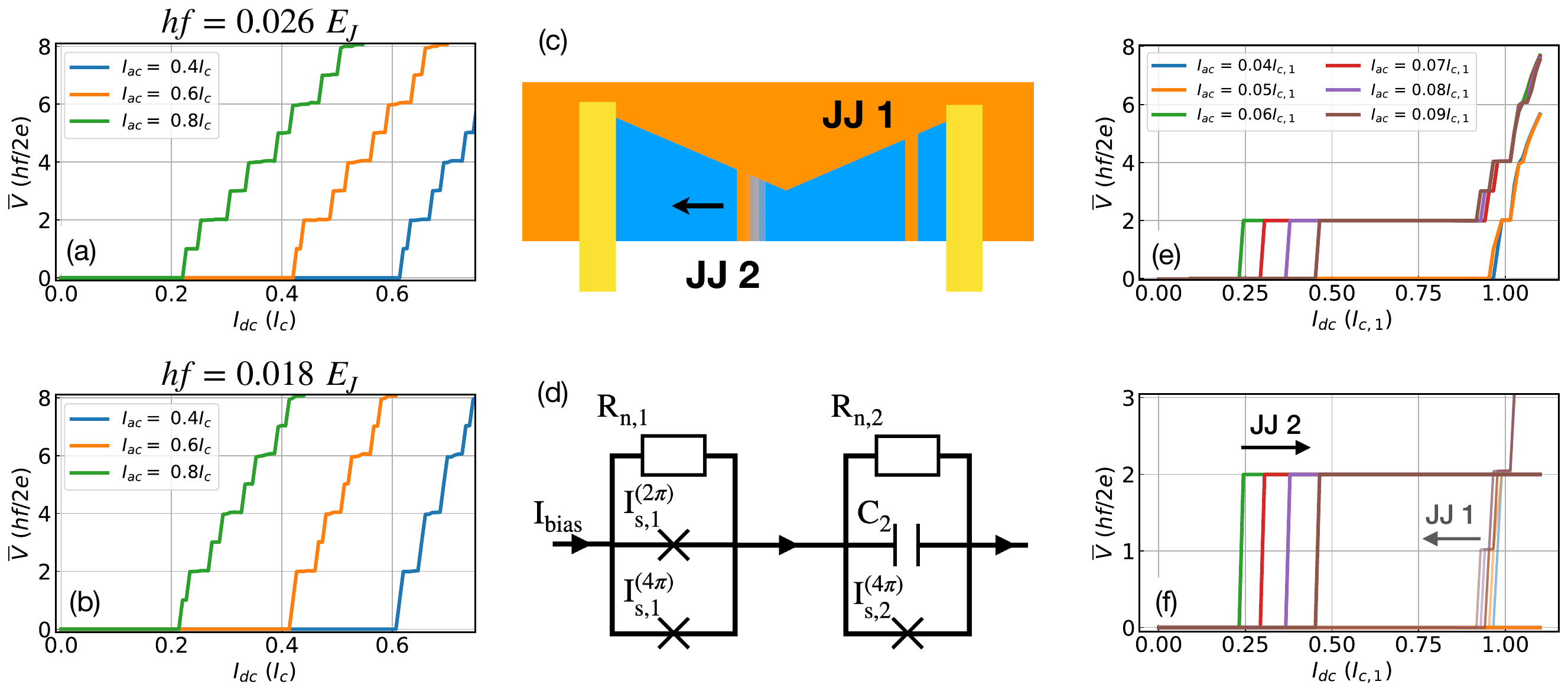} 
\vspace{-10pt}\caption{Shapiro steps calculated using the RCSJ model with LZTs using \textbf{a} $hf/E_J = 0.026$  and \textbf{b} $hf/E_J = 0.018$. The effective transparency for the modes undergoing LZTs was taken to be $\tau_{LZT} = 0.999$ and 
$I_{\tilde\tau}/I_{2\pi} = 2.0\%$.
\textbf{c}, Schematic of two PSLs in series (denoted JJ1 and JJ2) where JJ1 is fixed and JJ2 changes with an applied current bias. \textbf{d}, Circuit diagram for the dynamic two-junction model. \textbf{e}, Shapiro steps calculated using a two-junction model describing PSL motion. Here $\sigma = 6.7$, $I_{c,2}=I_{c,1} / 8$, $\alpha = 7$ for JJ2 and $hf = 0.09E_J$ in both junctions. \textbf{f}, 
Individual contributions of JJ1 and JJ2 to panel \textbf{e}.}
\label{Fig3-theory} \vspace{50pt} \end{figure}

Figure~\ref{Fig3-theory}~(a) shows the dependence of time-averaged voltage, $\overline{V}$, 
on the dc current for different values of the ac power when 
$\tilde{\tau}=0.999$, 
%
%
$I_{\tilde\tau}/I_{2\pi} = 2.0\%$,  
$E_J\equiv 2 e I_c R_N = 364.5~\mu eV$, 
and $hf=0.026 E_J$. 
This corresponds to a relatively high frequency regime compared to $f_{4\pi}$,
and we find that, for the powers considered, 
%
the Shapiro steps' structure does not exhibit missing steps, analogous to the experimental $V$--$I$ shown in Fig.~\ref{Fig3}~(b).
%
%
Figure~\ref{Fig3-theory}~(b) shows the results for the case when $hf=0.018 E_J$, all the other parameters being the same
as in Fig.~\ref{Fig3-theory}~(a). 
For this lower value of the frequency the contribution to the supercurrent from the high transparency channels
qualitatively affects the structure of the Shapiro steps: at low powers the odd steps are missing, as seen in 
the experimental results shown in Fig.~\ref{Fig3}~(f).
%
%


In the $dV/dI$ profile showed in Fig.~\ref{Fig3}~(g) we have two sets of peaks:
the ``standard" peaks outside the region where $dV/dI$ is mostly zero,
and isolated ``in-gap" peaks inside this region, present only when -9~dBm~$\lesssim P\lesssim$~-6~dBm and $|I|\lesssim$~0.15~mA.
To explain the presence of two sets of peaks in $dV/dI$ profiles like the one
shown in Fig.~\ref{Fig3}~(g) it is natural to assume that two PSLs in series are present.
%
%
One JJ, JJ1, with a large $I_c$ is responsible for the standard peaks, 
and one, JJ2, with a smaller $I_c$, is responsible for the in-gap peaks. 
The resulting effective circuit describing the dynamics of the two junctions is shown in  Fig.~\ref{Fig3-theory}~(d).
%
%

The $V$--$I$ characteristic associated to the in-gap peaks, see Fig.~\ref{Fig3}~(h),
has two very unique qualitative features:
(i)  
the critical current 
in the presence of ac bias ($I_{c,ac}$)
increases with the microwave power rather than decreasing, as expected for JJs;
(ii) the width of the second step is very large, larger than $I_{c,ac}$
and of the width of the conventional steps seen at higher powers.
The first feature strongly suggests that the critical current of the JJ responsible for the in-gap $dV/dI$ peaks might grow with the ac power. 
This can be understood by considering that a weak link created by a PSL can be affected by the biasing current: 
as the biasing current increases, 
if possible, the PSL will change to allow a larger supercurrent across the JJ. 
In our setup we can expect that, as the biasing current increases 
a PSL, initially at a point
close to the center of the ``bow-tie'',
might move away from the center and become wider, see Fig.~\ref{Fig3-theory}~(c),
causing JJ2 to have a larger $I_c$.
%

From the smallest value of $I_{c,ac}$ we estimate the minimum width of JJ2 to be approximately $50$~nm.
For such a small width we have that $R_N$ can be sufficiently large that even just one $4\pi$-periodic supercurrent channel
can be sufficient to have $f_{4\pi}\gtrsim$~1~GHz.
The fact that in the $V$--$I$ characteristic corresponding to the in-gap peaks shown in Fig.~\ref{Fig3}~(h) 
the absence of the first Shapiro step is very robust supports the hypothesis that its absence, at least for the smallest values of power and $I_{dc}$,
might be due to the topological nature of JJ2.
As discussed above, however, we cannot exclude contributions to the $4\pi$-periodic supercurrent arising from LZTs of highly transparent modes. 
For JJ2, a $4\pi$-periodic supercurrent channel appears to be sufficiently strong to determine the structure of the junction's Shapiro steps,
and so  for JJ2 we include only such a supercurrent channel.
We describe JJ1 via an RSJ model in which both a 2$\pi$- and 4$\pi$-periodic supercurrent channels are present. 
JJ2 is expected to form close to the middle of the bow-tie, a region where W is expected to be thinner
and so $I_c$ smaller. This suggests that for JJ2 $\sigma$ might not be very large and therefore that for JJ2
the capacitive term in Eq.~\ceq{eq:JJ} might not be negligible. Indeed, we find good agreement with the experimental results 
if for JJ2 we set $\sigma\sim 6-7$ and keep the capacitive term, resulting in the effective circuit model
shown in Fig.~\ref{Fig3-theory}~(d).
%
For the critical current of JJ2 we assume 
$
I_{c,2} = I_{c,2}^{(0)}+\alpha I_{ac} 
$
if $I_{dc}$ is smaller than $I_{\rm onset}$ and
$
I_{c,2} \approx I_{c,2}^{(0)}+\alpha I_{ac} + (I_{dc}-I_{\rm onset})
$
if $I_{dc}>I_{\rm onset}$, with $\alpha > 0$.
%
%
The extension of the width of the second Shapiro step in the $V$-$I$ characteristic of Fig.~\ref{Fig3}~(h)
allows us to fix the values of $I_{c,2}^{(0)}$, $I_{\rm onset}$, and $\alpha$ (see SI). 
Notice that given that we assume $I_{c,i}R_{N,i}=\pi\Delta/e=const.$, we have that for JJ2, as $I_{c,2}$ increases
$R_{N,2}$ decreases, which is reasonable if we attribute the increase of $I_{c,2}$ to an increase of the PSL's width.
Similarly, we keep the value of $\sigma$ fixed, implying that as $I_{c,2}$ increases
the capacitance also increase, consistent with the idea that 
the PSL moves to regions of the bow-tie with larger cross-sectional areas.
%
%

Fig.~\ref{Fig3-theory}~(e) shows the results for the $V$--$I$ characteristics, for different microwave powers, obtained integrating the RCJS model
corresponding to the circuit diagram shown in Fig.~\ref{Fig3-theory}~(d).
We see that we recover the main qualitative features observed experimentally at low frequencies and powers, Fig.~3~(h).
%
%
Figure~\ref{Fig3-theory}~(f) shows how the $V$--$I$ characteristic for JJ1 and JJ2 evolve as the microwave
power is increased: we see that $I_{c,ac}$ for the two junctions approach each other as $P$ increases. 
Given that the two JJs are in series, the full $V$--$I$ characteristic is given by the sum 
of the characteristics for JJ1 and JJ2.

\medskip
\medskip
\noindent
{\bf Discussion}
\smallskip

\noindent
%
%
In this work, by placing tungsten nanoislands on TI Bi$_{0.91}$Sb$_{0.09}$ using the focus ion beam technique,
we demonstrated a new approach to realize an air-stable heterostructure in which superconductivity is induced at the surface of a 3D TI.
By studying the transport properties in the dc limit 
we have shown that the system undergoes a Berezinskii-Koasterlitz-Thouless transition at $T=T_{BKT}\approx 3$~K.
%
%
We have shown that when the biasing current is larger than a threshold value, PSLs are formed, which can be described effectively as Josephson junctions.
We have estimated the length of the PSLs to be about 7~nm, and their width to be as small as 50~nm, 
making the geometry of the effective Josephson junction to be at the limit of current fabrication techniques. 
At low frequencies, the $V$--$I$ characteristic of PSL-formed JJ exhibits missing odd Shapiro steps. 
Our theoretical analysis suggests that for wide PSLs (of width of the order of a $\mu$m)
the absence of odd Shapiro steps is due to the presence of Andreev bound states with a large probability
to undergo a Landau-Zener transition when the phase difference across the PSL is close to $\pi$.
For PSLs of width $\sim 50$~nm we estimate the topological nature of the resulting Josephson junction
might be sufficient to explain the observed absence of odd Shapiro steps.
We showed how, by analyzing the response of the system to microwave radiation
it is possible to infer the presence of multiple PSLs and how the microwave power and 
dc current can affect their properties, in particular their width, and therefore the critical current of the 
effective Josepshon junction formed by the PSL.
Our results suggest that the width of a PSL can be controlled in a superconductor-TI microbridge with a bow-tie geometry
by tuning the biasing current, a result that complements approaches in which the 
PSL's nucleation site is controlled by other means, for instance the application of localized mechanical stress~\cite{Paradiso2019}.

The unique properties of the phase-slip lines in heterostructures like W-Bi$_{0.91}$Sb$_{0.09}$, and the 
possibility of engineering their width, make these structures a new platform to realize
topological Josephson junctions with geometries that stretch current fabrication techniques to the limit. 
The topological nature of such junctions could be further probed by measuring via tunneling contacts the
unique transport properties~\cite{Nichele2017, Huang2019} of the associated Majorana modes.
Replacing Bi$_{x}$Sb$_{1-x}$ with other TIs, e.g., (Bi$_{x}$Sb$_{1-x}$)$_2$Te$_3$, is also a promising step towards reducing the total number of conducting channels. 

\section{Acknowledgement}
We would like to thank Y. Rosen for helpful discussions, and K. Huang and A. A. Baker for assistance in performing the experiments. This work was performed under the auspices of the US Department of Energy by Lawrence Livermore National Laboratory under Contract No. DE-AC52-07NA27344. The project was supported by the Laboratory Directed Research and Development (LDRD) programs of LLNL (19-LW-040). J. J. Cuozzo and E. Rossi acknowledge support from DOE, Grant No DE-SC0022245. Sandia National Laboratories is a multimission laboratory managed and operated by National Technology and Engineering Solutions of Sandia LLC, a wholly owned subsidiary of Honeywell International Inc.~for the U.S.~DOE's National Nuclear Security Administration under contract DE-NA0003525. This paper describes objective technical results and analysis. Any subjective views or opinions that might be expressed in the paper do not necessarily represent the views of the U.S. DOE or the United States Government.

\medskip
\medskip
\noindent
{\bf Methods}
\medskip

\noindent Bi$_{0.91}$Sb$_{0.09}$ single crystals were synthesized by the modified Bridgman method with high purity (5N) Bi and Sb in a sealed quartz tube. The tube was heated up to 600 $^\circ$C for 1--2 days and shaken to homogenize the mixture. Then the tube was slowly cooled to 270 $^\circ$C over a period of 3.5 months. Finally the samples were annealed at 270 $^\circ$C for 3 days. Our devices are fabricated by pressing single crystal flakes onto a SiO$_2$/Si substrate with pre-fabricated Au electrodes. A micromanipulator is used to pick up the flake with a flat surface and move it to the center of the Au electrode pattern. Superconducting W-based focused-ion-beam technique was employed to perform the W deposition and tungsten hexacarbonyl W(CO)$_6$ gas was used as a precursor material. First, we deposited W leads with a thickness of $200$--$500$ nm by FIOB with a Ga$^+$ ion-beam current of 0.92 nA. Then, we deposited W pads to bridge the W leads to the pre-patterned Au electrodes. We iterated the W deposition process in combination with the transport measurements four times until realizing the zero-resistance state between the bottom W leads.  

Our transport measurements are carried out with a four-probe configuration to eliminate the contact resistance between W/Pt electrodes and Bi$_{1-x}$Sb$_x$. To attenuate electronic noise, $\pi$ filters are installed between the shielded cryostat and the measurement apparatus. For the Shapiro step measurements, microwave radiation is applied through a coaxial cable with a stripped end that is placed $1$--$2$ mm above the sample surface. All measurements are performed in a Helium-3 cryostat with a base temperature of 0.54 K.

\newpage

\bibliographystyle{alpha}

\end{document}